\documentclass[showpacs,superscriptaddress,prl,twocolumn]{revtex4}
\usepackage{amsmath,amssymb,amsfonts,amsthm}

\newtheorem{theorem}{Theorem}

\newcommand{\lc}{D\Phi}
\newcommand{\lca}{D\Phi^*}

\newcommand{\lb}{\mathcal{P}}

\newcommand{\lba}{\mathcal{P}^{*}}

\newcommand{\hkb}[2]{H^{#1}_{#2}}

\begin{document}
\title{A new geometric invariant on initial
  data for Einstein equations}

\author{Sergio Dain}
\email[E-mail: ]{dain@aei.mpg.de}
\affiliation{Albert-Einstein-Institut, am M\"uhlenberg
    1, D-14476, Golm, Germany.} 

\date{\today}
\begin{abstract}  
  For a given asymptotically flat initial data set for Einstein
  equations a new geometric invariant is constructed. This invariant
  measure the departure of the data set from the stationary regime,
  it vanishes if and only if the data is stationary. In vacuum, it can
  be interpreted as a measure of the total amount of radiation
  contained in the data.
\end{abstract}


\maketitle

\emph{Introduction.}  --- The mass of an asymptotically flat data for
Einstein equations measures the total amount of energy contained in
the spacetime. The mass is zero if and only the spacetime is flat.
However, the energy can be in different forms: it can be in an
stationary regime or in a dynamic one.  This difference is, of course,
physically important: gravitational radiation will be present only in
the second case. The purpose of this article is to construct a new
quantity that can measure how far the data are from the stationary
regime. In other words, this quantity measures how dynamic are the
data and it will be zero if and only if the data are stationary. In
vacuum, the dynamic is produced only by the gravitational field, then,
in this case, the quantity can be interpreted as a measure of the
total amount of radiation contained in the data.

The construction is based on a new definition of approximate
symmetries. An approximate symmetry satisfies an equation that has
always solution for a generic spacetime and the solution coincides
with the Killing vectors when the spacetime admits them. The new
quantity is related to the approximate timelike translation.

There are many physical situations which are dynamic but nevertheless
are expected to be described by data which are ``as close to
stationary data as possible'', this invariant provides a precise
meaning of this idea.  An important example is the
quasicircular orbit for binary black holes (see \cite{Cook00} and
reference therein).  The invariant constructed here will provide a
completely different way of finding this class of data and hopefully
will help to solve the discrepancies between the different current
approaches.

\emph{Symmetries, approximate symmetries and the constraint map.} ---
Let $S$ be a 3-dimensional manifold, and let
$\mathcal{M}_2,\mathcal{S}_2,\mathcal{X}, \mathcal{C}$ be the spaces of
Riemannian metrics, symmetric 2-tensors, vectors and scalar functions
on $S$ respectively. Let $h_{ab} \in \mathcal{M}_2$, $K_{ab}\in
\mathcal{S}_2$. The constraint map $\Phi: \mathcal{M}_2 \times
\mathcal{S}_2 \rightarrow \mathcal{C} \times \mathcal{X}$ is defined
as follows
\begin{equation}
  \label{eq:13c}
  \Phi  \begin{pmatrix}
 h_{ab}\\  
 K_{ab}
  \end{pmatrix}= \begin{pmatrix} R+K^2-K_{ab}K^{ab}\\  
 -D^bK_{ab}+D_a K
  \end{pmatrix}
\end{equation}
where $D_a$ is the covariant derivative
with respect to $h_{ab}$, $R=h^{ab}R_{ab}$, $R_{ab}$ is the
Ricci tensor of $h_{ab}$, $ K= h^{ab} K_{ab}$ and $a,b,c...$ denote
abstract indices, they are moved with the metric $h_{ab}$
and its inverse $h^{ab}$. The set $(S,h_{ab}, K_{ab})$ is called a
vacuum initial data set for Einstein equations if
$\Phi(h_{ab},K_{ab})=0$ on $S$.

We can compute the
linearization of $\Phi$ evaluated at $(h_{ab},K_{ab})$
\begin{equation}
  \label{eq:17b}
  \lc  \begin{pmatrix}
 \gamma_{ab}\\  
 Q_{ab}
  \end{pmatrix}= \begin{pmatrix} D^aD^b\gamma_{ab}-R_{ab}\gamma^{ab}-
    \Delta \gamma +  H  \\  
-D^bQ_{ab}+D_aQ - F_a
  \end{pmatrix} 
\end{equation}
and its formal adjoint
\begin{equation}
  \label{eq:17}
  \lca  \begin{pmatrix}
 \eta\\  
 X^a
  \end{pmatrix}= \begin{pmatrix} D_a D_b \eta -\eta R_{ab}-\Delta \eta
    h_{ab}+H_{ab}
   \\  
 D_{(a}X_{b)}-D^cX_ch_{ab} +F_{ab}
  \end{pmatrix} 
\end{equation}
where $\gamma=\gamma_{ab}h^{ab}$, $Q=Q_{ab}h^{ab}$ and $H$, $H_{ab}$,
$F_a$, $F_{ab}$ vanished when $K_{ab}=0$ (since
in this article we will only consider the time-symmetric case,
the explicit expression of these quantities will not be needed).   

The constraint map \eqref{eq:13c} is important not only because it
characterizes the initial data for Einstein equations, it also gives
the Hamiltonian of the theory (see \cite{Arnowitt62} and also
\cite{Regge:1974zd}, \cite{Beig87}). In particular, the adjoint map
\eqref{eq:17} gives the right hand side of the evolution equations in
the Hamiltonian formulation (see \cite{Fischer79} and reference
therein).  Moreover, $\lca$ has a remarkably property
\cite{MR50:15836}: the elements of the kernel of $\lca$ are the
\emph{symmetries} of the spacetime determined by the initial data
$(S,h_{ab},K_{ab})$. That is, if $(\eta,X^a)$ satisfies
$\lca(\eta,X^a)=0$ then the spacetime will have a Killing vector $\xi$
and $(\eta,X^a)$ are the projections of $\xi$ normal and tangential to
the space-like hypersurface $S$, respectively.

Motivated by this correspondence between the kernel of $\lca$ and
symmetries we introduce the concept of approximate symmetries. 
We will make use of the following related operator introduced by
Bartnik \cite{Bartnik04}
\begin{equation}
  \label{eq:19}
  \quad \lb  \begin{pmatrix}
 \gamma_{cd} \\  
 q_{pcd}
  \end{pmatrix}=\lc \begin{pmatrix} 
\gamma_{cd}
\\  
-D^pq_{pcd} 
  \end{pmatrix} 
\end{equation}
and its adjoint
\begin{equation}
  \label{eq:19b}
   \lba \begin{pmatrix}
 \eta \\  
 X_a
  \end{pmatrix}\equiv  \begin{pmatrix} 
1 & 0
\\  
0 & D_p 
  \end{pmatrix}\cdot \lca \begin{pmatrix}
 \eta \\  
 X_a
  \end{pmatrix},
\end{equation}
where the dot denotes matrix product.
 We say
that $(\eta,X^a)$ is an \emph{approximate symmetry} if it satisfies
the equation 
\begin{equation}
  \label{eq:25}
 \lb\lba(\eta,X^a)=0 
\end{equation}
and has the fall off behavior at
infinity of the Killing vectors in flat spacetime. Eq. \eqref{eq:25}
is solved for a given $(h_{ab}, K_{ab})$. Note that although the
composition $\lc\lca$ is formally well defined, there is a mismatch in
the units in the domains and ranges of these operators; that is why we
use $\lb$ instead of $\lc$ in Eq. \eqref{eq:25}.

This definition will be meaningful if: a) Every symmetry is also an
approximate symmetry b) For generic data which admit no symmetry there
exist always approximate symmetries c) We can uniquely associate each
approximate symmetry with a symmetry in flat space, that is, we have
approximate time translation, approximate boost, etc.  Since every
symmetry satisfies $\lca(\eta, X^a)=0$, a) trivially follows. In the
next section we will prove that b)-c) are true for time symmetric
initial data: in order to simplify the analysis in this article we
will assume that $K_{ab}=0$ and $X^a\rightarrow 0$ at infinity, that
is, we are only going to consider time translations and boosts (in the
flat case we have 4 solutions).  This is an important special case
since it includes all the relevant features and difficulties of the
equation.  The general case will be studied in a subsequent article.

Eq. \eqref{eq:25}  can be derived from a variational
principle, it is the 
Euler-Lagrange equation of the following functional
\begin{equation}
  \label{eq:22}
J(\eta,X^a)=\int_S \lba (\eta,X^a)\cdot \lba (\eta,X^a) d\mu,  
\end{equation}
where $d\mu$ is the volume element with respect to $h_{ab}$.  For a
symmetry we have $J=0$, for an approximate symmetry $J\geq 0$. If $J$,
evaluated at an approximate symmetry, is finite, we obtain new
invariants for the data, which measure how far is the approximate
symmetry from a symmetry.  This will be the case for the approximate
boosts. However, it turns out, that for the approximate timelike
translation $J$ diverges. This is because the approximate timelike
translation grows like $r$ at infinity and not like a constant as the
timelike translation. The new invariant (which will be denoted by
$\lambda$) is precisely the coefficient of $r$ in this expansion.

The variational principle \eqref{eq:22} is a generalization (which
includes the lapse function $\eta$ and it is of fourth order) of a
variational principle for $X^a$ studied in \cite{Smarr78a} which is
known as the minimal distortion gauge (see also the interesting
discussion in \cite{York79}). Finally, we want to point out that the
operators $\lc$ and $\lca$ has been recently used to construct new
kind of solutions for the constraint equations \cite{Corvino99}
\cite{Corvino:2003sp} \cite{Chrusciel:2003sr} \cite{Chrusciel:2002vb}.

\emph{The time symmetric case.} ---
If $K_{ab}=0$ and $X_a=o(1)$ then, contracting with $(\eta,X^a)$ Eq.
\eqref{eq:25} and integrating by parts, one concludes that $D_{(a}
X_{b)}=0$. Hence $X^a$ is a Killing vector of $h_{ab}$. But since
there are no Killing vectors which goes to zero at infinity (see
\cite{Christodoulou81}), we have $X^a=0$. Then, in this case, the
operators \eqref{eq:19}, \eqref{eq:19b} reduce to
\begin{align}
  \label{eq:5b}
  \lb(\gamma_{ab}) &= D_a D_b \gamma^{ab} - R_{ab}\gamma^{ab} -\Delta \gamma,\\
  \label{eq:3}
  \lba(\eta)_{ab} &= D_a D_b \eta -\eta R_{ab}-\Delta \eta h_{ab}.
\end{align}
Using  that $D^a\lba (\eta)_{ab}=-\eta D_b R /2$ and $h^{ab}\lba(\eta)_{ab}
=-2\Delta \eta -\eta R$ we obtain
\begin{multline}
  \label{eq:23}
   \lb \lba(\eta)=2\Delta\Delta \eta -D^a D^b\eta R_{ab}+ \eta R_{ab}R^{ab} +\\
 \frac{1}{2} \eta \Delta R +2 R \Delta \eta + \frac{3}{2} D^a \eta D_a
 R
\end{multline}
When $R=0$ the last three terms in \eqref{eq:23} vanish and
the operator $\lb \lba$ has a very simple expression.  For an
arbitrary domain $\Omega \subset S$ the boundary term is given by
\begin{multline}
  \label{eq:5}
  \int_{\Omega} \gamma^{ab} \lba(\eta)_{ab} \, d\mu = \int_{\Omega}
  \lb(\gamma_{ab}) \eta \, d\mu 
  + \\
 \oint_{\partial \Omega}\left(\eta D_a\gamma-\gamma D_a \eta+
   \gamma_{ab}D^b \eta - \eta D^b \gamma_{ab}  \right)  n^a \, ds,
\end{multline}
where $ds$ is the surface element with respect to $h_{ab}$, and $n^a$
is the unit normal of $\partial \Omega$ pointing in the outward
direction.

The fall off conditions on the fields can be conveniently written in
terms of weighted Sobolev spaces $\hkb{s}{\beta}$, where $s$ is a
non-negative integer and $\beta$ is a real number (see
\cite{Bartnik86} \cite{Cantor} \cite{Choquet81} and reference therein,
we will use the index notation of \cite{Bartnik86}).  We say that
$\eta \in\hkb{\infty}{\beta}$ if $\eta \in\hkb{s}{\beta}$ for all $s$.
The functions in $\hkb{\infty}{\beta}$ are smooth in $\mathbb{R}^3$
and have the fall-off at infinity $\partial^l \eta=o(r^{\beta-|l|})$.

For simplicity, we have not included matter fields in the constraint
map \eqref{eq:13c}. But then, if the metric is static and the topology
of the data is $\mathbb{R}^3$ it should be flat. In order to have
non-flat, vacuum, static metrics (i.e. Schwarzschild) we will allow
$S$ to have  $n$ \emph{asymptotic ends}, that is, for some
compact set $\Omega$ we have that $S\setminus \Omega =\sum_{k=1}^n
S_k$, where $S_k$ are open sets diffeomorphic to the complement of a
closed ball in $\mathbb{R}^3$.  Each set $S_k$ is called an
$\emph{end}$. We will also assume that $(S,h_{ab})$ is
\emph{asymptotically flat:} at each end $S_k$ there exists a
coordinate system $x_{(k)}^j$ such that we have in these coordinates
\begin{equation} 
\label{pf1}
h_{ij}-\delta_{ij}\in \hkb{\infty}{\tau}, \quad  \tau \leq -1/2
\end{equation}
where $\delta_{ij}$ denotes the flat metric and $i,j \cdots$, which
take values $1, 2, 3$, denote coordinates indices.  We will smoothly
extend the coordinate system $x_{(k)}^j$ to be \emph{zero} in
$S\setminus S_k$. 

We say that $\beta$ is \emph{exceptional} if $\beta$ is a non-positive
integer, and we say that $\beta$ is \emph{nonexceptional} if it is not
exceptional. The manifold $(S,h_{ab})$ will be called \emph{static} if
there exist a function $\nu \in H^4_{1/2}$ such that $\lba (\nu)_{ab}=0$.
Note that in this definition we allow $\nu$ to have zeros (horizons)
on $S$, as, for example, in Schwarzschild initial data.

We are interested in the kernel of $\lb\lba$ in
$\hkb{4}{\beta}$. Define
\begin{equation}
  \label{eq:12}
  N(\beta)=\dim \ker \left( \lb\lba: \hkb{4}{\beta} \rightarrow
    \hkb{0}{\beta-4} \right).
\end{equation}
The main result is given by the following theorem. 
\begin{theorem}
  Let $(S,h_{ab})$ be a complete, smooth, asymptotically flat,
  Riemannian manifold, with $n$ asymptotic ends.  Assume that  $\eta\in
  \hkb{4}{\beta}$  satisfies
  \begin{equation}
    \label{eq:27}
    \lb\lba \eta=0
  \end{equation}
  Then, $\eta\in\hkb{\infty}{\beta}$. Moreover, assume that $\beta$ is
  nonexceptional, then, we have the following:

(i) If  $\beta\leq 0$ then $N(\beta)=0$.

(ii) If $0<\beta\leq 1/2$ then $N(\beta)\leq 1$, and $N(\beta)= 1$ if
and only if $(S,h_{ab})$ is static. In this case we have $R=0$ and the
unique static solution $\nu$, $\lba \nu=0$, has at each end $S_k$ the
following fall off
\begin{equation}
  \label{eq:14}
  \nu-\nu^{(k)}_0 =o(r^{\tau}) \quad \partial^l\nu=o(r^{\tau-|l|})
\end{equation}
where $\nu^{(k)}_0 \neq 0$ are a constant and $\tau$ is
given by \eqref{pf1}.

(iii) If $1<\beta<2$ then $N(\beta)\geq 4n$.    At each end we have the
following four linear independent solutions  of Eq. \eqref{eq:27}
\begin{equation}
  \label{eq:1}
  \alpha_{(k)}=\lambda_{(k)} r_{(k)}+ \hat\alpha_{(k)}, \quad
  \eta_{(k)}^j=x_{(k)}^j + \hat\eta_{(k)}^j 
\end{equation}
with $\hat\alpha_{(k)} ,\hat\eta_{(k)}^j \in \hkb{4}{1/2}$. Where
$\lambda_{(k)}$ are
constants and $\lambda_{(k)}=0$ for some $k$  if and only if $(S,h_{ab})$ is static.

(iv) In the particular case $S=\mathbb{R}^3$ (which implies $n=1$)  we
obtain more information:

(ii') If $0< \beta <1$ then $N(\beta)\leq 1$ and $N(\beta)=1$ if and
only if $(S,h_{ab})$ is static and hence flat.

(iii') If $1<\beta <2$ then  $N(\beta)= 4$.

\end{theorem}

\begin{proof}
  The differential operator $\lb\lba$ is an elliptic operator of
  fourth order with smooth coefficients. Using the decay assumption on
  the metric \eqref{pf1} one can easily check that $\lb\lba$ is
  \emph{asymptotically homogeneous of degree} $m$ with $m\geq 4$ (for
  the definition of this concept see for example \cite{Cantor}, this
  is the standard assumption on the coefficients for elliptic operator
  on weighted Sobolev spaces, see also \cite{Choquet81}). Then, by the
  weighted Sobolev estimate \cite{Choquet81}, it follows that $\eta\in
  \hkb{s}{\beta}$ for every $s$.
  
  \emph{(i)--(ii)}. We use Eq. \eqref{eq:5} with
  $\gamma^{ab}=\lba{(\eta)}^{ab}$ and $\Omega=S$ to obtain
 \begin{equation}
   \label{eq:24}
  \int_{S}\lba(\eta)^{ab} \lba(\eta)_{ab} \, d\mu_h = \oint_{\partial
    S} B_a n^a \, ds,
 \end{equation}
with
\begin{equation}
  \label{eq:6}
  B_a=-2\eta D_a\Delta\eta +2\Delta \eta D_a \eta +
   \lba(\eta)_{ab}D^b \eta  -\frac{1}{2}\eta^2D_aR,
\end{equation}
where the boundary integral is performed in a two sphere at infinity
at every end. Since $\partial^l \eta=o(r^{\beta-|l|})$ we obtain
$B=o(r^{2\beta-3})$. If $\beta\leq 1/2$ then $2\beta-3\leq -2$ and the
boundary term vanished because $ds=O(r^2)$. We conclude that for
$\beta\leq 1/2$ we have $\lb\lba(\eta)=0 \iff \lba(\eta)=0$.  That is,
for $\beta\leq 1/2$ the kernel is not trivial if and only if the
metric is static.  To prove i) we use the result that there exists no
spacetime Killing vectors which go to zero at infinity
\cite{Beig:1996fk} \cite{Christodoulou81}.  The fall off behavior for
Killing vectors \eqref{eq:14} was proved in \cite{Beig:1996fk}.
Because the constants $v^{(k)}_0$ are non zero, this fall off implies
that $\nu$ is unique: assume that there exists another solution
$\nu'$, rescale $\nu'$ such that $\nu'-\nu =o(r^{\tau})$ at some end,
this contradict \eqref{eq:14}. Finally, we prove that static implies
$R=0$: in \cite{Corvino:2003sp} it has been proved that static implies
that $R$ is constant, by our falloff assumption it should be zero. 

\emph{(iii)} We use the Fredholm alternative in weighted Sobolev
spaces (see \cite{Cantor} and \cite{MR82j:35050}, note that we use
different conventions for the weights) to prove the existence of the
four independent solutions at each end: the equation $\lb\lba \eta =F$, with
$F\in\hkb{0}{\beta-4}$, will have a solution $\eta \in \hkb{4}{\beta}$
if and only if
\begin{equation}
  \label{eq:8}
  \int_S F\nu \, d\mu =0
\end{equation}
for all $\nu\in\hkb{0}{\beta'}$ such that $\lb\lba \nu
=0$, with $\beta'=1-\beta$.  

We prove first the existence of $\eta^j$ (in the following we will
suppress the end label $(k)$, all the calculations are done in one
arbitrary end). Set $\eta^j=x^j+\hat \eta^j$, then $\hat\eta^j$
satisfies the equation
\begin{equation}
  \label{eq:2}
   \lb\lba(\hat \eta^j)=- \lb\lba(x^j).
\end{equation}
We have that $\lb\lba(x^k)\in \hkb{0}{\tau -3}$. Since $\tau\leq
-1/2$ we can take $\hat \eta^j\in \hkb{4}{1/2}$ in Eq.  \eqref{eq:2}.
From the discussion above we have that a solution $\hat \eta^j\in
\hkb{4}{1/2}$ of \eqref{eq:2} will exist if and only if the right hand
side satisfies the condition \eqref{eq:8}. Since $\beta=\beta'=1/2$ in this
case, we can use ii) to conclude that a non trivial $\nu$ will exist
if and only if $(S,h_{ab})$ is static.  Then, if the metric is not
static we don't have any restriction and the solutions $\hat \eta^j$
exist. If the metric is static, we compute
\begin{align}
  \label{eq:10}
 \int_S \nu \lb\lba(x^j) &= \int_S  x^j \lb\lba(\nu)+\oint_{\partial
    S} B_a n^a \, ds,\\
&=  \oint_{\partial
    S} B_a n^a \, ds, \label{eq:10b} 
\end{align}
where
\begin{multline}
  \label{eq:7}
  B_a=-2\nu D_a\Delta x^j +2\Delta \nu D_a x^j + \lba(x^j)_{ab}D^b \nu \\
-2 x^j D_a\Delta \nu +2\Delta x^j D_a \nu,
\end{multline}
and we have used that $\lba(\nu)_{ab}=0$. 
We use  the  fall off \eqref{eq:14} for $\nu$, the fall off
\eqref{pf1} for the metric and the fact that $\Delta_\delta x^j=0$
(where $\Delta_\delta$ is the flat Laplacian) to conclude that
$B_a=o(r^{-2})$, then the boundary integral \eqref{eq:10b}
vanishes and  the solution exists also when the metric is static. 

For the solution $\alpha$ we proceed in an analogous way. Let
$\lambda\neq 0$ be 
an arbitrary constant. We have
that $\lb\lba(\lambda r)\in \hkb{0}{\tau -3}$ (here we use that
$\Delta_\delta \Delta_\delta r=0$).  If the
metric is not static there exists a solution $\hat \alpha\in \hkb{4}{1/2}$ of
\begin{equation}
  \label{eq:15}
 \lb\lba(\hat \alpha)=- \lb\lba(\lambda r).
\end{equation}
If the metric is static we can compute condition \eqref{eq:8} as we
did in in Eqs. \eqref{eq:10}--~\eqref{eq:7}
\begin{align}
  \label{eq:11}
 \int_S \nu \lb\lba(\lambda r) &=  -2\oint_{\partial S} \nu
 D_a\Delta (\lambda r)  n^a \, ds\\ 
 & =16\pi \lambda \nu_0, 
\end{align}
where $\nu_0\neq 0$ is the constant given in \eqref{eq:14} and we have
used $\Delta_\delta r =2/r$.  
Then, the constant $\lambda$ is zero if and only if the metric is
static.

For every end, we have constructed four independent solutions, then
 $N(\beta) \geq 4n$. 

\emph{(iv)} The Fredholm index of $\lb\lba$ is given by  
\begin{equation}
  \label{eq:16}
  \iota(\beta)=N(\beta)-N(1-\beta).
\end{equation}
When $S=\mathbb{R}^3$ the index $\iota(\beta)$ of $\lb\lba$ is equal
to the index $\iota_0(\beta)$ of the flat operator $\Delta_\delta
\Delta_\delta$ (see \cite{MR82j:35050}). To prove (ii')--(iii') we
will calculate $\iota_0(\beta)$ and use $\iota(\beta)=\iota_0(\beta)$.

Assume $\Delta_\delta
\Delta_\delta \nu =0$, that is $\Delta_\delta w=0$, $w=\Delta_\delta
\nu$. If $\nu \in \hkb{k}{\beta}$ with $\beta < 2$, then $w \in
\hkb{k-2}{\beta-2}$, we use the Liouville theorem proved in
Corollary 1.9 of \cite{Bartnik86}  to conclude that 
$w=0$. Then
$\Delta_\delta\nu=0$, using again this result we conclude that for
$0<\beta <1$ we have  $\nu=1$, and  for $1< \beta
<2$ we have $\nu = 1, x^j$.  
Then we have $\iota_0(\beta)=0$ for $0<\beta <1$
and $\iota_0(\beta)=4$ for $1<\beta <2$.

\emph{(ii')} We have $0=i_0(\beta)=i(\beta)=N(\beta)-N(\beta')$. 
We only need to prove the case $1/2<
\beta < 1$. In this case we have $0<\beta'<1/2$, using (ii) we get
$N(\beta')\leq 1$, and the equality holds if and only if the metric is
static.

\emph{(iii')} We have   $4=i_0(\beta)=i(\beta)=N(\beta)-N(\beta')$. 
and $\beta'<0$, using  (i) we get $N(\beta')=0$.  

\end{proof}

The new invariants are given by the constants $\lambda_{(k)}$. They
can be written as a boundary integral at each end $S_k$
\begin{equation}
  \label{eq:18}
  \lambda_{(k)}=\frac{-1}{8\pi}\oint_{\partial S_k} n^aD_a \Delta \alpha_{(k)} \, ds
\end{equation}
Note that $\alpha_{(k')}=o(r^{1/2})$ at $S_{k'}$, for $k'\neq k$, then
if we calculate \eqref{eq:18} for $\alpha_{(k')}$ we get zero.  Also,
if we compute this boundary  for $\eta^j$ we get zero.

If we assume $R=0$, using Eq. \eqref{eq:23} we get
another representation for $\lambda_{(k)}$ as a volume integral
\begin{equation}
  \label{eq:9}
  \lambda_{(k)}=\frac{1}{16\pi} \int_{S}  \alpha_{(k)}  R_{ab}R^{ab} \, d\mu.
\end{equation}

Since we have the freedom to rescale $\alpha$ by an arbitrary
constant, we need to normalize $\alpha$ if we want to compare
$\lambda$ for different data.  In the case of
$S=\mathbb{R}^3$ we have a natural normalization. Let $h_{ab}$ be a
metric on $\mathbb{R}^3$. Assume that there exists a smooth family
$h_{ab}(\epsilon)$ such that for $0 \leq \epsilon \leq 1$ it satisfies
the hypothesis of the previous theorem and $h_{ab}(1)=h_{ab}$,
$h_{ab}(0)=\delta_{ab}$. Then, for every $\epsilon$ we get a solution
$\alpha(\epsilon)$. The flat solutions $\alpha(0)$ are the constants.
We normalize $\alpha$ setting $\alpha(0)=1$. With this normalization,
we can calculate $\lambda$ up to second order in $\epsilon$
\begin{equation}
  \label{eq:4}
  \lambda \approx \frac{\epsilon^2}{16\pi}\int_S \dot R^{ab}\dot
  R_{ab} \, d\mu + O(\epsilon^3) 
\end{equation}
where $\dot R^{ab}=dR_{ab}(\epsilon)/d\epsilon|_{\epsilon=0}$.
 
The solutions $\eta^j$ provide a coordinate system near infinity. In
the non flat case this system is unique up to rotations, the
translation freedom at infinity is fixed because the constants are not
solutions of Eq. \eqref{eq:27}. 

Since $\lba(\eta^j)\in \hkb{0}{-3/2}$
we have that the functional $J(\eta^j)$ defined in \eqref{eq:22} is
finite. The solutions $\eta^j$ are the minimum of this functional with
the boundary conditions $\eta^j=x^j + o(r^{1/2})$ at infinity. The
numbers $J(\eta^j)$ are a measure of how far is the metric $h_{ab}$ of
having a boost Killing vector.  The fall of conditions of $\eta^j$ are
like the ones for the boost Killing vectors. In contrast, the solution
$\alpha$ has a different fall-off as the time translation.  This
difference is reflected also in the fact that $J(\alpha)$ is infinite,
it grows like $J(\alpha)\approx \lambda r$ at infinity.

\begin{acknowledgments}
  It is a pleasure to thank A. Ashtekar, R. Beig, H. Friedrich, G.
  Huisken and V.  Moncrief for illuminating discussions; special
  thanks to L. Szabados for pointing out the unit problem in the
  previous version and to P.T. Chru\'sciel for calling my attention to
  the operator $\lb$ in Ref. \cite{Bartnik04}.  This work has been
  supported by the Sonderforschungsbereich SFB/TR 7 of the Deutsche
  Forschungsgemeinschaft.
\end{acknowledgments}


\end{document}